\documentclass{vgtc}                          

\graphicspath{{figures/}{pictures/}{images/}{./}} 

\usepackage{times}                     

\usepackage{tabu}                      
\usepackage{booktabs}                  
\usepackage{lipsum}                    
\usepackage{mwe}                       
\usepackage[all]{nowidow}

\usepackage{mathptmx}                  

\vgtccategory{Theory}

\vgtcinsertpkg

\title{Data Humanism Decoded: A Characterization of its Principles to Bridge Data Visualization Researchers and Practitioners}

\author {Ibrahim Al-Hazwani\thanks{e-mail: alhazwani@ifi.uzh.ch}\\ %
        \parbox{1.4in}{\scriptsize \centering University of Zurich \\ Digital Society Initiative} %
\and Ke Er Zhang\thanks{e-mail: ke.zhang@uib.no}\\ %
     \scriptsize University of Bergen %
\and Laura Garrison\thanks{e-mail: laura.garrison@uib.no}\\ %
     \scriptsize University of Bergen %
\and Jürgen Bernard\thanks{e-mail: bernard@ifi.uzh.ch}\\ %
     \parbox{1.4in}{\scriptsize \centering University of Zurich \\ Digital Society Initiative}}

\abstract{
Data Humanism is a human-centered design approach that emphasizes the personal, contextual, and imperfect nature of data.
Despite its growing influence among practitioners, the 13 principles outlined in Giorgia Lupi's visual manifesto remain loosely defined in research contexts, creating a gap between design practice and systematic application.
Through a mixed-methods approach, including a systematic literature review, multimedia analysis, and expert interviews, we present a characterization of Data Humanism principles for visualization researchers.
Our characterization provides concrete definitions that maintain interpretive flexibility in operationalizing design choices.
We validate our work through direct consultation with Lupi.
Moreover, we leverage the characterization to decode a visualization work, mapping Data Humanism principles to specific visual design choices.
Our work creates a common language for human-centered visualization, bridging the gap between practice and research for future applications and evaluations.
} 

\keywords{Data Humanism, Critical Data Visualization, Human-Centered Visualization}

\begin{document}



\maketitle

\section{Introduction} 
Data Humanism has attracted increasing attention within the practice and research visualization community, as exemplified by projects like ``Data Garden''~\cite{offenwanger2024datagarden}, ``Data Selfie''~\cite{kim2019dataselfie}, and ``Dear Nature''~\cite{ferreira2024dear}. 
Perhaps the best-known exemplar is Giorgia Lupi's ``Dear Data'' project~\cite{lupi2016dear}, where she and Stefanie Posavec exchanged hand-drawn postcards visualizing personal data for a year. 
More recently, with her award-winning work \href{https://www.nytimes.com/interactive/2023/12/14/opinion/my-life-with-long-covid.html}{``1374 Days - My Journey with Long Covid''} for \textit{The New York Times}, Lupi demonstrated how personal, nuanced approaches to data representation can effectively communicate complex health experiences. 
These projects embody Data Humanism's core principles: emphasize the human aspects of data, embrace imperfection, and prioritize personal connection over standardization~\cite{lupi2017data}.
The growing popularity of Data Humanism introduces a more personalized and emotionally resonant approach to data experiences~\cite{segel2010narrative, hullman2011visualization}, presenting design considerations that challenge conventional practice.
\textit{Critical visualization} research has similarly gained momentum, challenging foundational assumptions in standard visualization approaches through frameworks such as feminist theory and critical theory~\cite{dork2013critical, d2023data}. 

Despite these interests, a considerable gap exists between design-driven examples like \href{https://www.behance.net/gallery/154534141/From-my-terrace}{From My Terrace}, \href{https://mappingdiversity.eu}{Mapping Diversity} or \href{http://www.storiesbehindaline.com}{Story Behind a Line} and research implementation guidance for those without formal design or humanist training. 
Computer scientists, researchers, and data practitioners often struggle to adapt these humanistic principles to their work~\cite{sprague2012exploring}. 
While Data Humanism's poetic nature inspires creativity, researchers need concrete guidance for systematic application, evaluation, and comparison across projects. 
Without clear definitions, researchers often struggle to consistently apply these principles or assess their effectiveness.
Additionally, the visualization research community increasingly values affect and engagement alongside traditional insight goals~\cite{bateman2010useful}, still frameworks and guidance for use are minimal~\cite{moere2011role}. 

Data Humanism's principles offer potential pathways to bridge these gaps, yet their characterization remains ill-defined. 
This raises an important question: \textit{``How can the Data Humanism principles be operationalized for the visualization community?''}

To address this question, we contribute by characterizing the 13 Data Humanism principles through a mixed-methods approach combining systematic literature review, multimedia analysis, and expert interviews. 
We validate our characterization through direct exchange with Lupi. Lastly, we demonstrate its practical application by decoding a visualization project. 
Our work bridges Data Humanism's design origins with visualization research, creating a common language for human-centered visualization that enables researchers and practitioners to operationalize these principles in future applications and evaluations.


\section{Data Humanism as Practice-Driven Lens to Critical Data Visualization}
\vspace{2mm}\textbf{Data Visualization Evolution --}
Data visualization has evolved multiple times to reflect changes in technological capabilities, societal needs, and epistemologies for reasoning about and representing data.
In his \textit{Tapestry 2018} keynote, Meeks~\cite{meeks2018} speaks of contemporary visualization approaches occurring in three waves.
The \textit{first wave} (1980s) focused on clarity and basic visualization types within the constraints of early software tools. 
This wave emphasized fundamental principles, including data-to-ink ratio and minimizing chart junk, and was heavily influenced by statistical approaches~\cite{tufte1985visual}.
The \textit{second wave} (1990s--2010s) was marked by structured approaches to visualization specifications and grammars~\cite{wilkinson2012grammar}. 
This period saw the emergence of tools and libraries~\cite{bostock2011d3} that enabled the systematic binding of data attributes to visual elements and interaction.
The \textit{third wave} (late 2010s) represents a synthesis of technical capability and human-centered approaches to emphasize long-term engagement, personalization, and integration within larger contexts. 
Data Humanism emerged within this third wave as a design-driven and practitioner-focused perspective on human-centered visualization~\cite{lupi2017data}. Data Humanism's concepts closely relate to theories and concepts in critical data visualization. 

\vspace{2mm}\textbf{Critical Data Visualization --}
Data visualization has experienced a critical turn in recent years, with scholars and practitioners increasingly questioning fundamental assumptions about data objectivity, neutrality, and the politics embedded in visualization practices. 
Johanna Drucker's concept of ``capta'', which posits that data are not merely given but are actively taken and interpreted, challenges the notion of data objectivity~\cite{drucker2015graphical}.
Similarly, the broader field of Critical Data Studies asserts that ``raw data is an oxymoron''~\cite{gitelman2013raw}, underscoring data as constructed and contextual~\cite{boyd2012critical}. We see a clear through-line of these assertions in Data Humanism through the principles \textit{imperfect} and \textit{subjective} data. 

Within visualization, Dörk et al.~\cite{dork2013critical} first formulated a set of ``Critical InfoVis'' design principles that explored the politics of visualization and power structures between visualization authors and their readers. 
Kennedy~\cite{kennedy2016work} further develops this critical perspective, analyzing how visualization conventions embed authority and objectivity through design choices, while Hill et al.~\cite{hill2016visualizing} explore how gender-, class-, and age-related judgments influence perceptions of visualizations.
Correll and Garrison~\cite{correll2024body}'s examination of historical anatomical illustrations explore the power dynamics and sociocultural norms encoded within scientific visualization practices. Such power structures and their entanglements have also been examined through the lens of feminist theories~\cite{d2023data}, which emphasize data, visualizations, and insights as inseparable from history, society, and the material world. 
We see these critical perspectives as the bedrock of the practitioner-centric principles of Data Humanism, in particular principles such as \textit{depict complexity} and \textit{data is people}. 
While sharing many critical concerns about objectivity and power, Data Humanism qualitative point of view and emphasis on artistic expression, imperfection, narrative, and data as design material offers practical, high-level guidance for visualization designers that are not immediately evident through theory. 
However, this guidance is non-specific, leaving many design choices open-ended and difficult to operationalize.  

The formal conceptualization of tacit knowledge has recently been tackled through an iceberg sensemaking process model~\cite{berret2024iceberg}. We are similarly interested in systematically characterizing the tacit knowledge behind Data Humanism. This motivates our work in characterizing Data Humanism's principles, enabling other researchers to operationalize these principles when developing or examining human-centered visualization.

\vspace{2mm}\textbf{Data Humanism --}
Proposed by Italian information designer Giorgia Lupi in 2017, Data Humanism synthesizes several existing ideas about human-centered design~\cite{kulyk2007human} and critical approaches to data visualization~\cite{loukissas2019all}.
It offers a \textbf{design- and practice-driven perspective} that reframes our conversations about data by advocating for engaging, personalized, design-driven data narratives that reconnect numbers to their underlying contexts, knowledge, and human experiences. 
Lupi's approach challenges the notion of data as objective and impersonal, or solely a source to be mined and exploited to augment human intelligence and insights.
The four overarching principles (\textit{Embrace complexity}, \textit{Move beyond standards}, \textit{Sneak context in}, and \textit{Remember that data is imperfect}) illustrate the core ideas of Data Humanism. 
In the accompanying visual manifesto\footnote{\tiny\url{http://giorgialupi.com/data-humanism-my-manifesto-for-a-new-data-wold}}, Lupi elaborates on 13 principles of traditional big data analytics with their humanistic counterparts. 
She aims to inspire practitioners to reimagine their relationship with data and foster a more thoughtful, nuanced, and human-centered visualization approach. 
To this end, the manifesto juxtaposes handwriting against a monospace, code-like font to reinforce its human-centered message. 
Despite its growing interest in visualization research, our community lacks a systematic characterization of its principles, seeing Data Humanism instead primarily as a call to action for alternative visualization design methods. We see value in a systematic characterization to enable targeted and widespread operationalization of these principles for the visualization community.

\section{Characterization of Data Humanism Principles} \label{sec:characterization}
Our work to bridge design practice and research applications of Data Humanism required a balanced approach. 
We wanted to maintain the interpretive flexibility valued by designers while providing concrete definitions useful to researchers. 
This section presents our method for characterizing the principles (first column in Table~\ref{tab:DaHu-final-characterizations}) and the resulting definitions (second column in Table~\ref{tab:DaHu-final-characterizations}), offering clear characterizations that visualization researchers and practitioners can apply in their work.

\begin{table*}[h]
\vspace{-5mm}
\resizebox{\textwidth}{!}{%
\begin{tabular}{lp{0.65\textwidth}ll}
\textbf{Data Humanism Principle} & \textbf{Characterization} & \textbf{Source(s)} & \textbf{Example}\\ \hline
Small data & \textit{Focus on individual stories within big data --} Prioritizes human-scale narratives across any dataset size, enriched with qualitative and contextual elements. Most relevant for narrative-driven projects and exploratory analysis. & \cite{richards2022questions} & \cite{kim2019dataselfie}\\
Data quality & \textit{Accept data as a lens, not a mirror --} Values quantitative accuracy alongside qualitative richness, with transparency in in data practices and how data reflects human complexity. Essential when building trust with audiences. & \cite{clever2023, quality2018} & \cite{panagiotidou2020data}\\
Imperfect data & \textit{Embrace flaws as a path to deeper meaning --} Recognizes data as shaped by human decisions and cultural contexts, accepting and leveraging imperfection as integral to interpretation. Valuable for revealing systemic biases and uncertainties.& \cite{data2023} & \cite{nowak2020designing}\\
Subjective data & \textit{Recognize multiple valid perspectives --} Views data as situated perspectives shaped by cultural lived experience. Critical for community-engaged projects and social justice contexts.& \cite{richards2022questions, woman2019} & \cite{correll2024body}\\
Inspiring data & \textit{Start conversations, not conclusions --} Treats data as an entry point for creative exploration that generates narratives and encourages critical reflection. Ideal for public engagement and educational contexts.& \cite{explore2020} & \cite{zhao2021chartstory}\\
Serendipitous data &  \textit{Allow for unexpected discoveries --} Embraces unpredictability (as in human experiences) by valuing unexpected connections rather than forcing predetermined patterns. Powerful for exploratory analysis and research.& \cite{lupi2017data} & \cite{jasim2023bridging}\\
Data possibilities & \textit{Experiment beyond conventions --} Promotes novel analysis, visualization, and interpretative approaches beyond conventional paradigms. Encourages innovation in research and artistic contexts it may require audience openness to unconventional formats.& \cite{lupi2017data} & \cite{bae2022making}\\
Data to depict complexity & \textit{Layer information for multiple audiences --} Balances complexity representation with accessibility through layered approaches allowing varied engagement levels. Effective when audiences can engage deeply.& \cite{complexity2014, lupi2014new} & \cite{windhager2024complexity}\\
Data drawing & \textit{Sketch to see and think through data --} Uses hand drawing to draft custom visual representations and a starting point to explore data in unique, tailored ways. Valuable for initial exploration and custom metaphor development.& \cite{lupi2015sketching, lupi2017data} & \cite{offenwanger2024datagarden}\\
Design-driven data & \textit{Design with people, not just for them --} Integrates design thinking throughout the data process, prioritizing human needs at each stage. Essential for participatory and community-centered projects it can require additional time and resources. & \cite{richards2022questions} & \cite{roberts2015sketching}\\
Spend time with data & \textit{Slow down the analytical process --} Encourages thoughtful engagement and deeper understanding while remaining sensitive to audience context. Applicable in exploratory research and reflection-oriented projects.& \cite{lupi2014new, lupi2017data} & \cite{odom2019investigating}\\
Data is people & \textit{Remember the humans behind every point --} Recall data as a representation of real lives and experiences, requiring ethical consideration throughout.Critical for any human-subjects data and fundamental for maintaining ethical standards in sensitive contexts. & \cite{lupi2017data, richards2022questions} & \cite{mittenentzwei2023disease}\\
Data will make us more human & \textit{Use data to connect, not just to compute --} Envisions ethical data use that deepens human understanding, empathy, and connection. Essential for public-facing communication projects.& \cite{lupi2017data} & \cite{elli2022visualizing}
\end{tabular}
}
\vspace{1mm}
\caption{Characterizing Data Humanism: conceptual foundations and example visualizations. \textit{Sources} shows key references used to define the principles; \textit{Examples} points to visualizations that the principle aligns with.}
\label{tab:DaHu-final-characterizations}
\vspace{-5mm}
\end{table*}

\subsection{Methodology} \label{sec:methdology}
We used a mixed-methods approach with four distinct phases to characterize Data Humanism principles.
\textbf{Ideation:} We began with a comprehensive review of Data Humanism works from academic literature and multimedia sources, including podcasts and blog posts. 
This phase allowed us to ideate and form an initial characterization of the 13 Data Humanism principles.
\textbf{Refinement:} We then conducted semi-structured interviews with data visualization researchers and practitioners to gather insights into their interpretation of these principles and feedback on our characterization.
\textbf{Validation:} We validated our interpretations with Lupi to ensure our characterization accurately represented the conceptual underpinnings of her visual manifesto. 
\textbf{Case Study}: Finally, we leveraged our characterization to decode whether and which Data Humanism principles are applied in a visualization project.
The results are presented in Section~\ref{sec:analysis}.

\vspace{2mm}\textbf{Ideation --}
We started by examining visualization research publications citing the manifesto to establish a foundation for Data Humanism principles. 
Given limited academic coverage, our research expanded to include a broader media content analysis~\cite{braun2016collecting}.
This allowed us to explore Lupi's perspectives through various formats, including podcasts, magazine articles, and blog posts, where she discusses themes related to the manifesto.
For podcast content, we listened to and transcribed key segments addressing the Data Humanism manifesto, its principles, and Lupi's underlying design philosophy. 
For written media, we collected articles explicitly focused on Data Humanism, multi-layer storytelling, and materials from Lupi's former design studio, \textit{Accurat}. 
At the end of the ideation phase, we had a working set of the Data Humanism principles characterizations. 

\vspace{2mm}\textbf{Refinement --}
To refine our characterization, we conducted nine semi-structured interviews with data visualization researchers and practitioners who have used or explored Data Humanism in their work. 
This deliberate mix ensured diverse perspectives representing theoretical and applied data visualization approaches. It recognizes that an idea, once released to the world, may take on a life of its own in the hands of different data designers.  
The interview protocol focused on gathering feedback on our preliminary characterization of the Data Humanism principles.
Of the nine participants, four could not join oral interviews and participated through an adapted protocol for email responses. 
This protocol maintained the original structure and qualitative elements and allowed us to collect their insights in written form. 
This phase provided a pluralistic and nuanced characterization of each principle sourced from research, media, and visualization experts with lived experience engaging with Data Humanism. 

\vspace{2mm}\textbf{Validation --}
We validated our interpretations with Lupi to ensure our characterization accurately represented the conceptual underpinnings of her visual manifesto. 
Through email exchanges, we sought clarification on specific terminology and confirmation about our characterization of the principles for visualization researchers.

\vspace{2mm}\textbf{Case Study --}
To demonstrate how Data Humanism principles can be effectively combined in practice, we analyze 'Tied in Knots'~\cite{elli2022visualizing}. 
We selected this visualization project because it explicitly incorporates multiple Data Humanism principles while addressing a sensitive social issue.
Through this illustrative case study, we deconstruct this project into key elements, mapping each to our characterized Data Humanism principles.

\subsection{Analysis} \label{sec:analysis}
We present our characterization of Data Humanism principles in Table~\ref{tab:DaHu-final-characterizations}, developed through an iterative process synthesizing insights from systematic reviews and expert interviews, and validated with Giorgia Lupi. 
Each principle is supported by foundational sources that shaped our understanding, alongside exemplar visualizations demonstrating these principles in practice. 
The complete evolution from initial to final characterizations is documented on OSF\footnote{\tiny\url{https://osf.io/tmnra/?view_only=06edc12deed64382a354dac3a7a62049}}.
Our work provides the visualization community with a characterization for discussing human-centered approaches, enabling researchers to explicitly position their work within Data Humanism.

\vspace{2mm}\textbf{Ideation --}
Our reviews identified 50 works referencing Data Humanism across academic papers (35), books (4), theses (3), and other sources (8). While frequently appearing alongside ideas like Data Feminism~\cite{d2023data} (65\% of citations) and in digital humanities contexts (31\%), most citations (48\%) reference Data Humanism in introductory and conclusion sections without active implementation, revealing a critical gap in operationalizing the manifesto.
While applications focused on participatory design and social justice show promise~\cite{ferreira2023interactions}, the lack of clearly defined principles motivated our work to provide a detailed characterization of Data Humanism to bridge theory and implementation. 

Our media analysis of 18 podcasts and four articles (2014--2024) traced Data Humanism's evolution from a design philosophy to a visual manifesto challenging automated, big data-oriented visualization approaches. Our analysis extracted key features of the manifesto, detailed below, that supported our characterization of the individual principles. 
Lupi describes Data Humanism as ``connecting numbers to what they stand for: our imperfect, messy human lives," challenging data objectivity by noting that ``even sensor data reflects \textbf{human choices about what to measure and ignore}''~\cite{clever2023}. 
She emphasizes \textbf{manual processes}, particularly sketching, for a deeper engagement with data and critical questioning of their assumptions. 
Her \textbf{analog approach} explores ``categories of data...possible correlations...to shape the story,'' demonstrating how physical engagement enhances understanding~\cite{complexity2014}. 
Lupi's methodology employs \textbf{multi-layered storytelling} with manual exploration and drawing~\cite{lupi2014new,lupi2015sketching}.
Her former studio, \textit{Accurat}, describes a distinctive Data Humanist strategy for visualizing data that preserves data complexity while ensuring accessibility through thoughtful design~\cite{Accurat_2016}. 
Their process begins with manual sketches to understand data structure, develops custom visual metaphors tailored to each dataset, and creates multi-layered visualizations balancing immediate narratives with deeper exploratory details.

\vspace{1mm}\textbf{Refinement --}
Participants in our refinement phase came from visualization, computer science, biomedical sciences, architecture, and information design, with experience ranging from theoretical research to commercial applications.
They reported high familiarity with Data Humanism on a 5-point Likert scale (M = 4.54, SD = 0.634).
Their feedback revealed consensus and constructive critique, leading to a more comprehensive characterization. While all 13 principles received input, six emerged as particularly challenging and were discussed in greater depth. The others required minimal revision, with general agreement on their initial framing.

The \textit{small data} principle generated significant discussion. 
Multiple participants (P01, P09, P10, P13) questioned the apparent dichotomy between small and big data approaches. 
P09 and P13 emphasized that thoughtful data handling, rather than dataset size alone, determines whether visualizations embody humanistic values. 
P13 noted that small data's value lies in its transparency and ``situatedness'', while P09, P10, and P13 connected small data approaches with qualitative research methodologies. This led us to frame the principle to emphasize thoughtful handling, transparency, and situatedness rather than as a size-based opposition.

Regarding \textit{data quality}, participants (P09, P11, P13) argued that quality considerations should extend beyond qualitative data. 
P13 emphasized that quality should be measured by how accurately data reflects reality, regardless of data type. 
Moreover, P10 and P13 highlighted the intrinsic connection between quality and transparency in data representation. 
We expanded our characterization to include all data types and their accuracy in reflecting reality.

The \textit{subjective data} principle prompted several suggestions for refinement. 
P08 recommended explicitly referencing ``positionality statements'' and ``cultural lived experience'' in its characterization. 
P13 proposed reframing subjectivity as ``situated perspectives of reality'' rather than filtered representations. 
Multiple participants (P07, P08, P10) emphasized the crucial role of context in shaping both data collection and interpretation processes. We refined our characterization to acknowledge data as embodied viewpoints shaped by context, experience, and cultural understanding.

While participants broadly supported the \textit{data to depict complexity} principle, they suggested important nuances. 
P08 noted this principle should not be universally applied, while others (P06, P12) stressed balancing complexity with accessibility to ensure visualizations remain approachable. We refined our characterization to recognize that complexity should be balanced with accessibility.

The \textit{data drawing} principle was well received, with P12 highlighting its pedagogical value.
Some participants (P08, P13) suggested this principle relates more to the process and relationship with data than literal hand-drawing. 
P07 emphasized the value of sketching during data exploration and avoiding limitations imposed by software tools. We broadened our characterization to focus on creative, exploratory engagement with data through hand-drawing.

Perspectives on \textit{spending time with data} varied. 
While P09, P10, and P12 strongly supported this principle, P08 questioned whether ``slowness'' should be a defining characteristic. 
P12 suggested incorporating interaction as an essential aspect of engagement, and multiple participants (P08, P09, P12) noted its context-dependent nature. 
Our characterization emphasizes meaningful engagement and interaction while recognizing context-dependent application to reflect these points.
We include participants' refinement suggestions in our final characterization in Table~\ref{tab:DaHu-final-characterizations}.

\vspace{2mm}\textbf{Validation --}
Lupi expressed general agreement with our characterization, noting that our work effectively captured the fundamental essence of Data Humanism and her manifesto.
During our exchange, we clarified two terms that generated discussion among our interviewees: `small` and `serendipitous`.
The term `small` in \textit{small data} extends beyond personal-level data. In Lupi's view, \textit{small data} encompasses ``any data that transcends cold aggregated numbers, incorporating granular stories enriched with qualitative and anecdotal elements.'' This broader interpretation reinforces Data Humanism's emphasis on narrative and contextual richness.
Regarding the term `serendipitous,` Lupi explained that this concept stems from a fundamental observation about human experience: ``There is hardly anything predictable in life.'' 
This terminology underscores Data Humanism's embrace of uncertainty and complexity in data representation, positioning visualization as an exploratory rather than purely predictive endeavor.

These insights from Lupi helped us validate our final characterization, ensuring that it reflects the design philosophy's core tenets while acknowledging the intentionally interpretive nature.


\vspace{2mm}\textbf{Case Study --}
'Tied in Knots'~\cite{elli2022visualizing} expresses individual narratives as distinct ``knot'' structures that preserve the context of a given experience, embracing the \textit{small data} principle. Its spatial layout and interaction design encourage thoughtful navigation and organic discovery of human stories, exemplifying the \textit{spend time with data} and \textit{serendipitous data} principles. By crafting animations that progressively reveal the knot structures of the creators' harassment metaphor, it fulfills \textit{data to depict complexity}. By integrating multimodal testimony, including text and audio components, it facilitates an empathetic connection, embracing \textit{subjective data}, \textit{data possibilities}, and \textit{data will make us more human}.

\section{Discussion and Future Work}
Our characterization bridges design-driven approaches with visualization research methodologies, complementing scholarly critical theories with practical implementation insights. 
A key tension emerged between systematic frameworks favored by researchers and the interpretive flexibility valued by design practitioners—a potential conflict acknowledged by interviewees and Lupi herself. 
Rather than providing rigid rules, we propose using our characterization as a guiding definition for operationalizing design choices while maintaining methodological flexibility.

Reflecting on existing visualization research through the lens of our characterization reveals how Data Humanism principles are already influencing the field, both explicitly and implicitly (Example column in Table~\ref{tab:DaHu-final-characterizations}).
For instance, we can see how Nowak et al.~\cite{nowak2020designing}'s research on avalanche decision-making tools resonates with the \textit{imperfect data} principle by treating imperfection not as noise to eliminate but as a meaningful signal requiring thoughtful design—effectively transforming a perceived limitation into a design opportunity. 
Similarly, Kim et al.~\cite{kim2019dataselfie}'s DataSelfie system embodies the \textit{small data} principle by prioritizing qualitative dimensions over aggregation, demonstrating a pathway toward more personalized visualization tools. 
In Windhager~\cite{windhager2024complexity}'s work, we can recognize elements of the \textit{data to depict complexity} principle as complexity is treated as a design material to be strategically distributed rather than minimized. 

These reflections point toward promising research directions: developing heuristic evaluation metrics that capture qualitative goals beyond efficiency and accuracy; applying Data Humanism to explainable AI through contextual understanding; and creating educational materials that teach data literacy through personal data collection. These possibilities suggest pathways toward human-centered visualizations that reconcile systematic approaches with contextual sensitivity.

\section{Conclusion}
We present the first comprehensive characterization of Data Humanism principles, bridging a gap between design practice and visualization research. 
Our mixed-methods approach helps us characterize all 13 Data Humanism principles, connecting Lupi's design-driven approach with visualization research methodologies and critical perspectives. In doing so, we introduce a common language for researchers and practitioners to discuss human-centered visualization strategies. We demonstrate how Data Humanism principles influence visualization research, revealing an implicit alignment between practitioner-led innovation and academic exploration. 
Our characterization opens new research directions for developing and evaluating visualizations beyond traditional efficiency metrics, incorporating qualitative goals like engagement, personal connection, and ethical information representation. 

\acknowledgments{
We thank the University of Zurich and the Digital Society Initiative for supporting the research. Moreover, we thank Mennatallah El-Assady for her insighful discussions and feedback.}

\bibliographystyle{abbrv-doi}

\bibliography{reference}
\end{document}